\documentclass{elsart}
\usepackage{amssymb}
\usepackage{epsfig,epsf,graphicx,amsmath}
\begin{document}
\begin{frontmatter}
\title{Constraints on two-neutron separation energy in the Borromean $^{22}$C nucleus}
\author[IFT]{M. T. Yamashita}\footnote{Email: yamashita@ift.unesp.br},
\author[UNIFESP]{R. S. Marques de Carvalho},
\author[ITA]{T. Frederico} and
\author[UFF,IFT]{Lauro Tomio}
\address[IFT]{Instituto de F\'\i sica Te\'orica, UNESP - Univ Estadual Paulista,\\
C.P. 70532-2, CEP 01156-970, S\~ao Paulo, SP, Brazil}
\address[UNIFESP]{Universidade Federal de S\~ao Paulo - UNIFESP, \\
04039-002, S\~ao Paulo, SP, Brazil.}
\address[ITA]{Departamento de F\'\i sica, Instituto Tecnol\'ogico de
Aeron\'autica, CTA,\\
12228-900, S\~ao Jos\'e dos Campos, Brazil.}
\address[UFF]{Instituto de F\'{\i}sica, Universidade Federal
Fluminense,\\ 24210-346, Niter\'oi, RJ, Brazil.}
\date{\today}
\maketitle
\begin{abstract}
The recently extracted matter radius of carbon isotope $^{22}$C
allows us to estimate the mean-square distance of a halo neutron with
respect to the center-of-mass of this nucleus. By considering this
information, we suggest an energy region for an experimental
investigation of the unbound $^{21}$C virtual state.
Our analysis, in a renormalized zero-ranged three-body model,
also indicates that the two-neutron separation energy in
$^{22}$C is expected to be found below $\sim$0.4~MeV, where
the $^{22}$C is approximated by a Borromean configuration with
a pointlike $^{20}$C and two $s$-wave halo neutrons.
A virtual-state energy of $^{21}$C close to zero,
would make the $^{22}$C, within Borromean nuclei
configurations, the most promising candidate to present an
excited bound Efimov state or a continuum three-body resonance.
\newline\newline
PACS {27.30.+t,21.10.Gv, 21.45.+v,11.80.Jy, 21.10.Dr}
\end{abstract}
\end{frontmatter}

\section{Introduction}

In the breakthrough experiment reported by K. Tanaka et al.~\cite{tanaka10},
the matter radius of the carbon isotope $^{22}$C was extracted via a
finite-range Glauber analysis under an optical-limit
approximation of the reaction cross section, for $^{22}$C on a
liquid hydrogen target, measured around 40 MeV/nucleon.
The extracted matter radius presents a huge
value of 5.4$\pm$0.9~fm (for a viewpoint on \cite{tanaka10}, see also
Ref.~\cite{kemper10}.), which characterizes this nucleus as the heaviest
halo nuclei discovered until now. For the two-neutron separation energy,
$S_{2n}$, they also quote a value of 0.42$\pm$0.94~MeV. These experimental
results, together with other well-known properties of carbon
isotopes~\cite{jensen1,amorim,brown2001,revs}, indicate that $^{22}$C is
weakly-bound, having a very large two-neutron halo with the $^{20}$C as a core,
such that the corresponding observables are probably dominated by the
tail of the three-body wave function in an ideal $s$-wave three-body model,
as considered in Ref.~\cite{suzuki}. In addition, within a $n-n-^{20}$C
configuration, we have the $^{22}$C as a Borromean halo system, considering
that a neutron ($n$) and $^{20}$C is known as an unbound system.

In view of its very low-energy properties, within the $n-n-$core halo-nuclei systems,
the nucleus $^{20}$C has been cited previously~\cite{jensen1,amorim,prl-mazumdar,prl-yama,plb}
as a good candidate to present three-body Efimov states~\cite{Efimov}.
Considering that this nucleus is more compact than the $^{22}$C, with its ground state
in a probable $(0d_{5/2})^6$ configuration, it is also quite natural to suggest
the $^{22}$C as being still more favorable to have an Efimov state~\cite{suzuki}, with
its halo predominantly produced by a $(1s_{1/2})^2$ component.
However, to infer about the possibility of existence of Efimov excited states
in an ideal $s$-wave two-neutron halo nucleus like $^{22}$C~\cite{suzuki} is
crucial to have a measurement of the virtual state energy of $^{21}$C.

The characteristics of $^{22}$C, roughly described in the first and second
paragraphs, allow us to use a Dirac$-\delta$ (zero-range) interaction, as
reviewed in Refs.~\cite{revs,tomio-epj}, acting on $s$-wave to study this problem.
In the zero-range limit three scales emerge for describing the full long-range
structure of the $n-n-^{20}$C wave function: the virtual $n-n$ energy, the $s$-wave
virtual state energy of the neutron in $^{21}$C and the two-neutron separation
$S_{2n}$. The information on the unbound $n-^{20}$C virtual energy is unknown
and $S_{2n}$ has an uncertainty that is about twice its own value.

In this study, we calculate a region to guide the experiments to search for
the $^{22}$C two-neutron separation energy. By first considering that the
virtual state energy of $^{21}$C is varying from 0 to 100 keV, and that the bound-state
energy of $^{22}$C is given in an interval from 100 to 1500 keV, we calculate the
mean-square distance of the halo neutron to the center-of-mass (CM) of the
corresponding three-body system as a function of  $S_{2n}$. Then, by using
the extracted one-nucleon mean-distance $r_n$ and its uncertainties
as constraints, we are able to estimate a reasonable region for the search
of the two-neutron separation energy in $^{22}$C, as well as the corresponding
region of the virtual state energy of $^{21}$C (directly related to a negative
scattering length of the $n-^{20}$C system).

\section{Neutron-neutron-$^{20}C$ model}

The three-body halo wave function allows us to calculate the neutron mean-square 
distance to the corresponding three-body CM. This model has already been applied 
with success to describe halo radii in Ref.~\cite{praios} and two-neutron 
correlation functions in Ref.~\cite{corr}.

The available quantity that can be used to define limits on the
two-neutron halo $^{22}$C binding is the extracted matter
radius, $R_M^{^{22}{\rm C}}=5.4\pm0.9$ fm, that was recently
reported in Ref.~\cite{tanaka10}. The root-mean-square distance,
from the CM of the $n-n-^{20}$C to one of its halo neutrons, can be
estimated by using the additional information on the matter
radius of the loosely bound $^{20}$C, which is given in
Ref.~\cite{ozawa} ($R_M^{^{20}{\rm C}}= 2.98\pm$0.05~fm). In view of
the large difference between the radius of $^{22}$C and $^{20}$C, we
consider it is a reasonable approximation to assume $^{20}$C as the
core for the present purpose, such that we still can use a
three-body approach. The result of our estimation is given by the
following:
\begin{eqnarray}
r_n \simeq \sqrt{\frac{22}{2}\left[(R_M^{^{22}{\rm C}})^2-
\left(\frac{20}{22}R_M^{^{20}{\rm C}}\right)^2\right]}\approx 15\pm 4\; {\rm fm} \ ,
\label{rn}
\end{eqnarray}
where $r_n\equiv\sqrt{\langle r_n^2\rangle}$ and  $R_M^{^{i}{\rm C}}\equiv
\sqrt{\langle (R_M^{^{i}{\rm C}})^2\rangle}$, with $i=20,22$. This simple approximation
shows that $^{22}$C is the largest known halo along the neutron dripline. By using this
value, we will be able to define a region where the $^{21}$C virtual energy can be found,
as well as the corresponding two-neutron separation energy, $S_{2n}$, in $^{22}$C.

\subsection{Subtracted Faddeev Equations}

In the following, the Faddeev formalism is developed by considering a renormalized
zero-range three-body model for a system with a core ($c$), which will be the $^{20}{\rm C}$
in the present work, and two-identical particles (the neutrons). The mass of the core is given by
$m_c=A m_n$, where $A$ defines the mass ratio and $m_n$ is the neutron mass. Throughout
this article, we will use units such that $\hbar=m_n=1$. In the renormalization procedure,
the kernel regularization is done via a subtraction method also considered in \cite{praios}.
After partial wave projection, the $s$-wave coupled subtracted integral equations, for two
neutrons and a core, can be written in momentum space by a coupled equations for
the spectator functions $\chi_c(x)\equiv \phi_c(x)/x$ and  $\chi_n(x)\equiv \phi_n(x)/x$,
as follows:
\begin{eqnarray}
\phi_{c}(y)&=&2\tau_{nn}(y;\epsilon_3)\int_0^\infty dx\;
G_1(y,x;\epsilon_3)\phi_{n}(x),
\label{chi1} \\
\phi_{n}(y)&=&\tau_{nc}(y;\epsilon_3)
\int_0^\infty dx \;
\left[G_1(x,y;\epsilon_3) \phi_{c}(x)
+A G_2(y,x;\epsilon_3) \phi_{n}(x)\right] ,
\label{chi2}
\end{eqnarray}
where
\begin{eqnarray}
\tau_{nn}(y;\epsilon_3)&\equiv &\frac{1}{\pi}
\left[\sqrt{\epsilon_3+\frac{A+2}{4A} y^2} +
\sqrt{\epsilon_{nn}} \right]^{-1},  \label{tau1}
\\
\tau_{nc}(y;\epsilon_3)&\equiv
&\frac{1}{\pi}\left(\frac{A+1}{2A}\right)^{3/2}
\left[\sqrt{\epsilon_3+\frac{A+2}{2(A+1)} y^2} +
\sqrt{\epsilon_{nc}}\right]^{-1}\ , \label{tau2}
\\
G_1(y,x;\epsilon_3)&\equiv &\log
\frac{2A(\epsilon_3+x^2+xy)+y^2(A+1)}{2A(\epsilon_3+x^2-xy)+y^2(A+1)}
\nonumber \\
&-& \log\frac{2A(1+x^2+xy)+y^2(A+1)}{2A(1+x^2-xy)+y^2(A+1)} ,
\label{G1} \\
G_2(y,x;\epsilon_3)&\equiv & \log \frac{2(A\epsilon_3
+xy)+(y^2+x^2)(A+1)}{2(A \epsilon_3-xy)+(y^2+x^2)(A+1)} \nonumber
\\ &-& \log \frac{2(A +xy)+(y^2+x^2)(A+1)}{2(A -xy)+(y^2+x^2)(A+1)}.
\label{G2}
\end{eqnarray}
In the above, we are using the odd-man-out notation for the spectator
functions $\chi$. The indexes $n$ or $c$ in $\chi$ indicates
the spectator particle. The momentum and energy variables are
written in terms of a momentum three-body scale $\mu_{(3)}$, which is
used in our subtractive regularization procedure to renormalize the
originally singular Faddeev equations. The units considered in
Eqs.~(\ref{chi1}-\ref{G2}) are such that all quantities are dimensionless.
In view of that, the corresponding dimensionless energies for the three-body
system are given by $\epsilon_3\equiv S_{2n}/\mu_{(3)}^2,$ $\epsilon_{nn}
\equiv -E_{nn}/\mu^2_{(3)},$ $\epsilon_{nc} \equiv -E_{nc}/\mu^2_{(3)}$,
where $E_{nn}=-143$ keV and $E_{nc}$ are, respectively, the $n-n$ and the
$n-^{20}$C virtual-state energies.

\subsection{The form factor and the mean-square radius}

The mean-square distance of the neutron to the CM of the three-body system
is calculated from the derivative of the Fourier transform of the
respective matter density with respect to the square of the momentum
transfer. The Fourier transform of the one-body densities defines the
respective form factor, $F_n(q^2)$, as a function of the dimensionless
momentum transfer $\vec q$. Thus, for the mean-square distance of the
neutron to the CM of $^{22}$C, we have~\cite{praios}
\begin{eqnarray}
\langle r^2_n\rangle= -6\left(\frac{21}{22}\right)^2
\frac{dF_n(q^2)}{dq^2}\bigg|_{q^2=0}, \label{ra}
\end{eqnarray}
where the form factor is defined as:
\begin{eqnarray}
F_n(q^2)&=&\int d^3p\;d^3k\; \Psi_{n}\left(\vec{p}+\frac{\vec{q}}{2},\vec{k}\right)
\Psi_{n}\left(\vec{p}-\frac{\vec{q}}{2},\vec{k}\right) \ .
\label{F3}
\end{eqnarray}

The above three-body wave function, $\Psi_n$, in momentum space are given in
terms of the spectator functions $\chi$ as:
{\small
\begin{eqnarray}
\label{psi} &&\Psi_{n}(\vec p,\vec k)=
\left(\frac{1}{S_{2n}+\frac{A+1}{2A} \vec k^2+
\frac{A+2}{2(A+1)}\vec p^2}-\frac{1} {\mu_3^2+\frac{A+1}{2A} \vec k^2+
\frac{A+2}{2(A+1)} \vec p^2}\right) \\ \nonumber
&&\times\left[\chi_{c}\left(\left|\vec z - \frac{A \vec
y}{A+1}\right|\right)+\chi_{n}\left(\left|\vec y\right|\right) +
\chi_{n}\left(\left|\vec z+\frac{\vec y}{A+1}\right|\right)\right],
\end{eqnarray}
}
where $\vec k\equiv \vec z\mu_3$ is the relative momentum of the pair
and $\vec p\equiv \vec y\mu_3$ is the relative momentum of the spectator particle
to the pair.

\section{Results and Conclusion}

The calculation of the neutron average distance to the CM of
$^{22}$C demands as input the $S_{2n}$, the energies of the virtual
$s$-wave states of the $n-n$ and $^{21}$C systems. The
unbound $^{21}$C virtual state is poorly known. In our model
we assumed small values of this virtual state between 0-100 keV. The
one-neutron mean distance to the CM,
$r_n\equiv\sqrt{\langle r^2_n \rangle}$, derived from Eqs. (\ref{ra})
and (\ref{F3}) and using the wave
function (\ref{psi}) can be written as a general function
${\mathcal R}_n$, dependent on the two-body energies, as:
\begin{eqnarray}
r_n=
{\mathcal R}_n\left(\pm\sqrt{\epsilon_{nn}\mu_{(3)}^2},\pm\sqrt{\epsilon_{nc}\mu_{(3)}^2}\right) \ ,
\label{rscal}
\end{eqnarray}
where the plus sign (minus) refers to bound (virtual) two-body
subsystem. The value of the separation energy is given by
$\epsilon_3=S_{2n}/\mu_{(3)}^2$. To convert all results of the calculations to
the physical units we have to introduce the physical value of $S_{2n}$ in (\ref{rscal}).
In this case the value of the parameters $\epsilon_{nn}$ and
$\epsilon_{nc}$ are determined as:
\begin{eqnarray}
\epsilon_{nn}=-\frac{E_{nn}}{\mu_{(3)}^2}=-\frac{E_{nn}}{S_{2n}}\epsilon_3
\ {\text { and }} \epsilon_{nc}=-\frac{E_{nc}}{S_{2n}}\epsilon_3 \ .
\label{e2}
\end{eqnarray}
From (\ref{rscal}) and (\ref{e2}) , the average distance from the neutron to the CM of the
system is given by
\begin{eqnarray}
r_n=\frac{1}{\sqrt{S_{2n}}}
{\mathcal R}_n\left(
-\sqrt{\frac{|E_{nn}|}{S_{2n}}\epsilon_3},-\sqrt{\frac{|E_{nc}|}
{S_{2n}}\epsilon_3}
\right) \ . \label{rscal1}
\end{eqnarray}
The limit cycle~\cite{mohr} is achieved when $\epsilon_{nn}$ and
$\epsilon_{nc}$ tends to zero and it is used to compute the radius
of the shallowest $n-n-c$ bound state. Therefore, in this limit, the
dependence on $\epsilon_3$ can be dropped out:
\begin{eqnarray}
r_n=\frac{1}{\sqrt{S_{2n}}} {\mathcal R}_n\left(
-\sqrt{\frac{|E_{nn}|}{S_{2n}} },-\sqrt{\frac{|E_{nc}|}{S_{2n}} }
\right) \ . \label{rscal2}
\end{eqnarray}
In practice such limit is achieved fast and the first cycle is
enough for the application we are considering (see Ref.~\cite{yama02}).

\begin{figure}[tbh!]
\centerline{\epsfig{figure=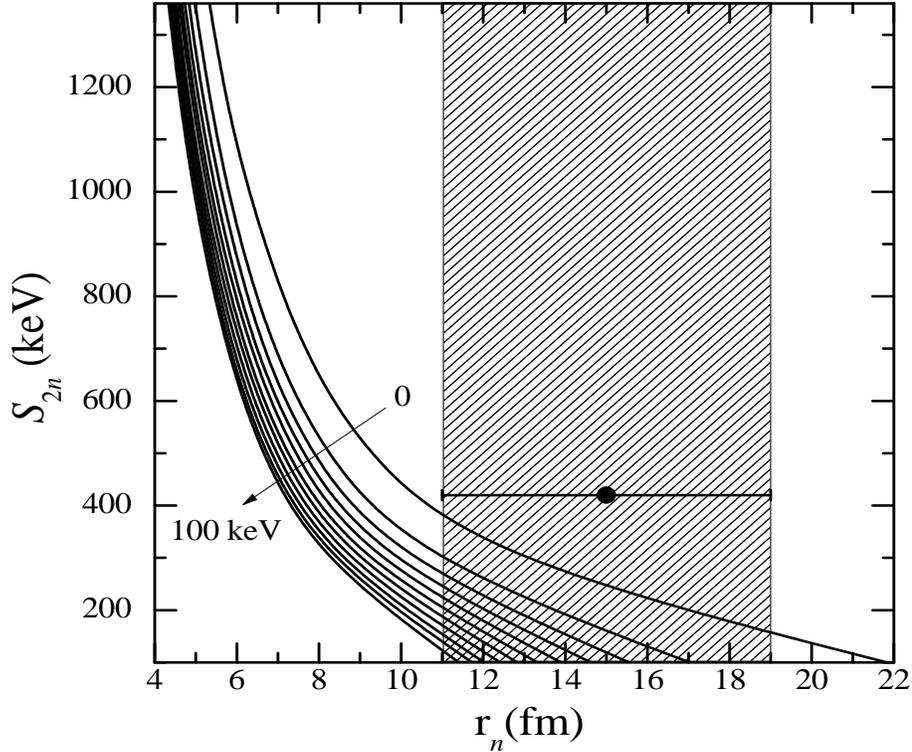,width=12cm,height=10cm}}
\caption{Two halo neutron separation energies in $^{22}$C ($S_{2n}$)
are given in terms of root-mean-square distances of a halo neutron with
respect to the three-body CM  ($r_n$). Each curve
is calculated for a given $^{21}$C virtual-state energy, varying
in steps of 10 keV, from 0 to 100 keV (indicated by the arrow). The
shaded area, involving the experimental point, corresponds to the
region defined by  100 keV $\le S_{2n}\le$ 1360 keV, with 11 fm
$\le r_n\le$ 19 fm.} \label{c22fig}
\end{figure}

From experimental data, we have
$r_n =15\pm 4$~fm, as given by Eq.~(\ref{rn}), and the singlet $n-n$
virtual state energy $E_{nn}=-143$~keV. Therefore, in order to use the model
results from (\ref{rscal2}), we have to assume values for the unknown virtual
state energy of $^{21}$C, to be able to get some information on $S_{2n}$ in
$^{22}C$. In Fig.\ref{c22fig}, we display our results for the
separation energy for different values of the $s$-wave neutron virtual
state in $^{21}C$, ranging from 0 up to 100 keV. The experimental
values of $S^{(exp)}_{2n}=0.42\pm 0.94$~MeV \cite{tanaka10} and
$r_n=15\pm4$~fm are shown in the figure.

We observe that, for a given $S_{2n}$, the $r_n$ decreases as the
absolute value of the virtual state energy increases. This can be
explained as follows: as the virtual state energy increases, the
interaction between the neutron and the core becomes weaker.
Therefore, one can fix a given three-body energy by decreasing the
size of the system~\cite{praios}. By taking into account the value
of $15\pm4$~fm, one obtains $S_{2n}$ below $\sim$0.4 MeV for a
neutron in $^{21}$C bound at the threshold. This result is not far
from the central experimental value of 0.42~MeV. We note that a
small increase in the virtual state energy up to 20~keV, drops the
upper limit of $S_{2n}$ to $\sim$0.3~MeV. Indeed, the finite-range
Glauber analysis under an optical-limit approximation of the
reaction cross section, for $^{22}$C on a liquid hydrogen target,
measured around 40 MeV/nucleon~\cite{tanaka10}, indicates that the
observed large enhancement of the cross-section compared to the
neighbor carbon isotopes, suggests that values of $S_{2n}$ below
0.4~MeV would be possible.

The three-body approximation that we have considered for $^{22}$C,
where  the $^{20}$C is the core, can be justified by comparing the
size of $^{20}$C with the mean distance of the halo neutrons of
$^{22}$C and also considering that the halo neutrons in $^{20}$C are
bound with about 3.5 MeV, one order of magnitude greater than
$S_{2n}$ in $^{22}$C. Thus, the halo neutrons in $^{22}$C have a
much larger probability to experience the long-range $1/r^2$
potential derived by Efimov than in $^{20}$C, as the corresponding
wave function tail is extending far beyond the size of $^{20}$C.
Therefore, the Efimov physics should be much more evident in the
properties of $^{22}$C ground state than in the corresponding
properties of $^{20}$C.

In a microscopic 5-body description, beyond the present model, the
four neutrons out of $^{18}$C, should be in a fully antisymmetric
wave function due to the proposed separation of scales. As the
$s$-wave radial wave functions corresponding to the neutrons in the
halo of $^{20}$C and in $^{22}$C have  different sizes, an
antisymmetric wave function can be built. If all spectator neutron
interactions are dominated by only $s$-waves, as in our model, the
Pauli exclusion principle would make the halo neutrons in $^{22}$C
much less bound than in $^{20}$C, which indeed seems to be the case.
In essence, with the above remarks, we should emphasize that our
three-body model for $^{22}$C is not excluding a three-body model
for $^{20}$C as having a two-neutron halo or an Efimov state for
$^{20}$C very near the scattering threshold~\cite{amorim}.

One possible correction to our results is due to the interaction
range. Range corrections in the calculation of different mean
distances were performed by Canham and Hammer~\cite{canham}. By
taking the $^{11}$Li (Borromean $n-n-^9$Li system) for comparison,
where $S_{2n}\sim$ 300 keV and the neutron average distance to the CM
is around 6 fm, the correction is a fraction of 1 fm ~\cite{canham}
for a fixed $S_{2n}$, $^{10}$Li virtual state energy and $nn$
scattering length. Therefore, we also expect corrections of the same
magnitude, or even smaller, considering that the core is larger but
the average distance of the neutron to the CM is more than twice. We
should stress that effects from the detailed core dynamics in our
calculation are implicitly carried out by the three-body energy,
which in our framework is an external parameter.

Summarizing, from the extracted matter radius of $^{22}$C, by
using a renormalized three-body zero-range model, we estimate the
mean-square distance of a halo neutron with respect to the CM of the
$^{22}$C. From such estimate, we suggest a possible region for an
experimental search of both $S_{2n}$ of $^{22}$C and the $n-^{20}$C
virtual state energy. The $^{22}$C is approximated by a Borromean
three-body system composed by a point-like core of $^{20}$C and two
$s$-wave halo neutrons. The validity of our findings relates to the
assumption of a large halo compared to the typical range of the
nuclear interaction. We are confident that the guidance provided by
this work would help the search for the $^{22}$C energy from an
experimental analysis.

Finally, based on Fig. 1 of Ref. \cite{amorim}, where it is calculated a
region for the appearance of excited Efimov states, we would like
to mention that a $^{21}$C with an energy close to zero can make the $^{22}$C
as the most promising Borromean candidate to present excited Efimov states,
or a continuum resonance~\cite{resonance}.

\section*{Acknowledgments}
Our thanks to the Brazilian agencies FAPESP and CNPq for partial financial support.
We also would like to thank Prof. Yasuyuki Suzuki for suggesting the study of
$^{22}$C.

\end{document}